\documentclass{article}

\usepackage{arxiv}

\usepackage[utf8]{inputenc} % allow utf-8 input
\usepackage[T1]{fontenc}    % use 8-bit T1 fonts
\usepackage{url}            % simple URL typesetting
\usepackage{booktabs}       % professional-quality tables
\usepackage{amsfonts}       % blackboard math symbols
\usepackage{nicefrac}       % compact symbols for 1/2, etc.
\usepackage{microtype}     
\usepackage{amsmath}
\usepackage{amssymb}
\usepackage{graphicx}% Include figure files
\usepackage{dcolumn}% Align table columns on decimal point
\usepackage{bm}% bold math
\usepackage{xcolor}

\title{Identification of Anomalous Diffusion Sources by Unsupervised Learning}

\author{
  Raviteja Vangara\\
  Theoretical Division\\ Los Alamos National Laboratory\\ Los Alamos NM, USA\\
  \texttt{rvangara@lanl.gov} \\
   \And
   Kim \O. Rasmussen\\
     Theoretical Division\\ Los Alamos National Laboratory\\ Los Alamos NM, USA\\
  \texttt{kor@lanl.gov}\\
  \And
  Dimiter N. Petsev\\
   Dept of Chemical and Biological Engineering\\
   Center for Microengineered Materials\\University of New Mexico, NM, USA\\
  \texttt{dimiter@lanl.gov}\\
  \And
Golan Bel$^*$\\
Dept of Solar Energy and Environmental Physics\\
Ben-Gurion University of the Negev\\Sede Boqer Campus 8499000, Israel\\
\texttt{bel@bgu.ac.il}
\And
Boian S. Alexandrov\thanks{Corresponding authors}\\
Theoretical Division\\ Los Alamos National Laboratory\\ Los Alamos NM, USA\\
  \texttt{boian@lanl.gov} \\

}

\begin{document}
\maketitle

\begin{abstract}
Fractional Brownian motion (fBm) is a ubiquitous diffusion process in which the memory effects of the stochastic transport result in the mean squared particle displacement following a power law, $\langle {\Delta r}^2 \rangle \sim t^{\alpha}$, where the diffusion exponent $\alpha$ characterizes whether the transport is subdiffusive, ($\alpha<1$), diffusive ($\alpha = 1$), or superdiffusive, ($\alpha >1$).
Due to the abundance of fBm processes in nature, significant efforts have been devoted to the identification and characterization of fBm sources in various phenomena. In practice, the identification of the fBm sources often relies on solving a complex and ill-posed inverse problem based on limited observed data. In the general case, the detected signals are formed by an unknown number of release sources, located at different locations and with different strengths, that act simultaneously. This means that the observed data is composed of mixtures of releases from an unknown number of sources,
which makes the traditional inverse modeling approaches unreliable. Here, we report an unsupervised learning method, based on Nonnegative Matrix Factorization, that enables the identification of the unknown number of release sources as well the anomalous diffusion characteristics based on limited observed data and the general form of the corresponding fBm Green's function. We show that our method performs accurately for different types of sources and configurations with a predetermined number of sources with specific characteristics and introduced noise.
\end{abstract}

% keywords can be removed

\section{\label{sec:level1}Introduction}

Anomalous diffusion has been observed in numerous systems, and a variety of underlying mechanisms have been discussed \cite{Bouchaud_1990,Metzler_2000}.
Anomalous diffusion is associated with the nonlinear dependence of the mean square displacement on time, $\langle \Delta {\bf{r}}(t)\rangle\sim t^\alpha$ with $\alpha\neq1$.
Different mechanisms can lead to the same asymptotic dependence of the mean square displacement on time, or alternatively, to the same diffusion exponent. However, such processes can differ in their propagator, probability distribution function \cite{Bouchaud_1990,Metzler_2000}, aging \cite{Barkai_2003}, and ergodic properties \cite{Bel_2005}. Fractional Brownian motion (fBm) is a common model for anomalous diffusion which stems from long range correlations, stationarity and scaling of the increments \cite{Mccauley_2007}. The fBm has applications in many fields including finance, climate, solar activity, hydrology, turbulence and many others \cite{Biagini_2008}.
The fBm diffusion exponent, $\alpha$,  determines the diffusion regime. When $\alpha<1$,  the process is subdiffusive;  when $\alpha>1$, the process is superdiffusive; and when $\alpha=1$, we have the normal Brownian diffusion \cite{Bouchaud_1990}.
Examples of subdiffusion are the kinetics of passive molecular  tracers in lipid bilayers \cite{kneller2011communication}, hopping transport in disordered systems \cite{lucioni2011observation},  propagation of nonlinear waves in quasiperiodic potentials \cite{larcher2012subdiffusion}, evolution of index prices in financial systems \cite{staliunas2003anticorrelations}, and many others.  Superdiffusion is typically associated with active processes, and has been observed in living cells \cite{caspi2000enhanced}, in radiative transport \cite{pereira2004photon}, in intracellular particle motion \cite{reverey2015superdiffusion}, in transport processes in porous media \cite{Molz_1993,Molz_1997,Cortis_2004}, and in many other cases.

Due to the non-Markovian nature of fBm, much effort has been devoted to the development of inference methods for the diffusion characteristics from observed data. Most of this research has focused on parameter inference using inverse modeling \cite{Coeurjolly_2001,Cheng_2009,Wei_2010,Yaozhong_2011,Furati_2014,Jin_2015,Ning_2015,Janno_2018,Zhang_2018,Cai_2019}. Inverse modeling typically requires identification of the characteristics of the sources as well as of the medium. For fractional diffusion, there is the additional complexity of the propagator describing the process, which has focused the inverse modeling efforts on one-dimensional problems \cite{Liu_1999,Cheng_2009,Wei_2010,Furati_2014,Jin_2015,Janno_2018,Zhang_2018}, while less attention has been devoted to two- and three-dimensional problems \cite{Zhang_2007,Ning_2015}. Furthermore, most research in this field assumes that the measurements exist at high spatial and temporal resolutions, a situation that is rarely encountered. Usually, the identification of release sources relies on solving a complex ill-posed inverse model against limited amounts of observed data. Importantly, in most of the naturally occurring fBm phenomena, the number of the release sources (and their locations and strengths) is unknown, which severely limits the reliability of the classical inverse modeling.

Recently, a hybrid method based on Matrix Factorization (NMF) combined with Brownian diffusion Green's function, called hNMF, was proposed for  identifying the properties of unknown number of emission sources releasing simultaneously \cite{Boiandiff}.

Here, we report a generalization of hNMF that enables identification of fBm diffusion processes (sub, super or normal diffusion) based on limited observed data and fBm Green's functions. To demonstrate the performance of our generalized method, we generate examples that include anisotropic two-dimensional fBm with drift and a predetermined number of release sources. We show that our method determines accurately the unknown number, locations, and properties of the release sources used to generate the data (including point sources as well as spatially and temporally extended sources). We also demonstrate that the generalized hNMF correctly determines the generalized diffusion coefficients, the advection velocity, and the diffusion exponent, $\alpha$, and that it accurately estimates the spatial and temporal extension of the emission in the presence of noise.

\section{Problem Formulation}\label{sec;Form}
We focus on the problem of identifying the anomalous diffusion process characteristics based on a limited number of concentration time series recorded by spatially fixed detectors. This is fundamentally an inverse problem, but unlike many previous works, our method is based on the recording of the concentration at specific locations rather than on single particle tracking. The general problem is complex due to the large number of parameters and the need to identify several different characteristics.
%what is uniquely different?
The fundamental step is to be able to identify the unknown number of sources and their locations and strengths. This is not a simple task due to the fact that the concentrations, recorded by each detector, potentially contain contributions from all sources. Further, each source may have a finite extent in space and may have a temporally varying amplitude of emission. In addition to the number of sources, $\tilde{N}_s$, their locations, $x_s, y_s$, and strengths, $q_s$ (where $s =1, 2, ...,\tilde{N}_s$) the medium characteristics must also be determined.
The medium, in which the fBm process takes place, has the following characteristics: 1) the type of diffusion, i.e., super, sub or normal diffusion, which is estimated by the value of the diffusion exponent, $\alpha$ (see subsection \ref{subs:fbm}); 2) the generalized diffusion coefficients, $D_x$ and $D_y$, which for the general anisotropic case, can be different in the two ($x$ and $y$) directions; and 3) the drift velocity, ${\vec{u}}$, which for transport in porous media, affects the diffusion.

\begin{figure*}[ht]
\includegraphics[width=\linewidth]{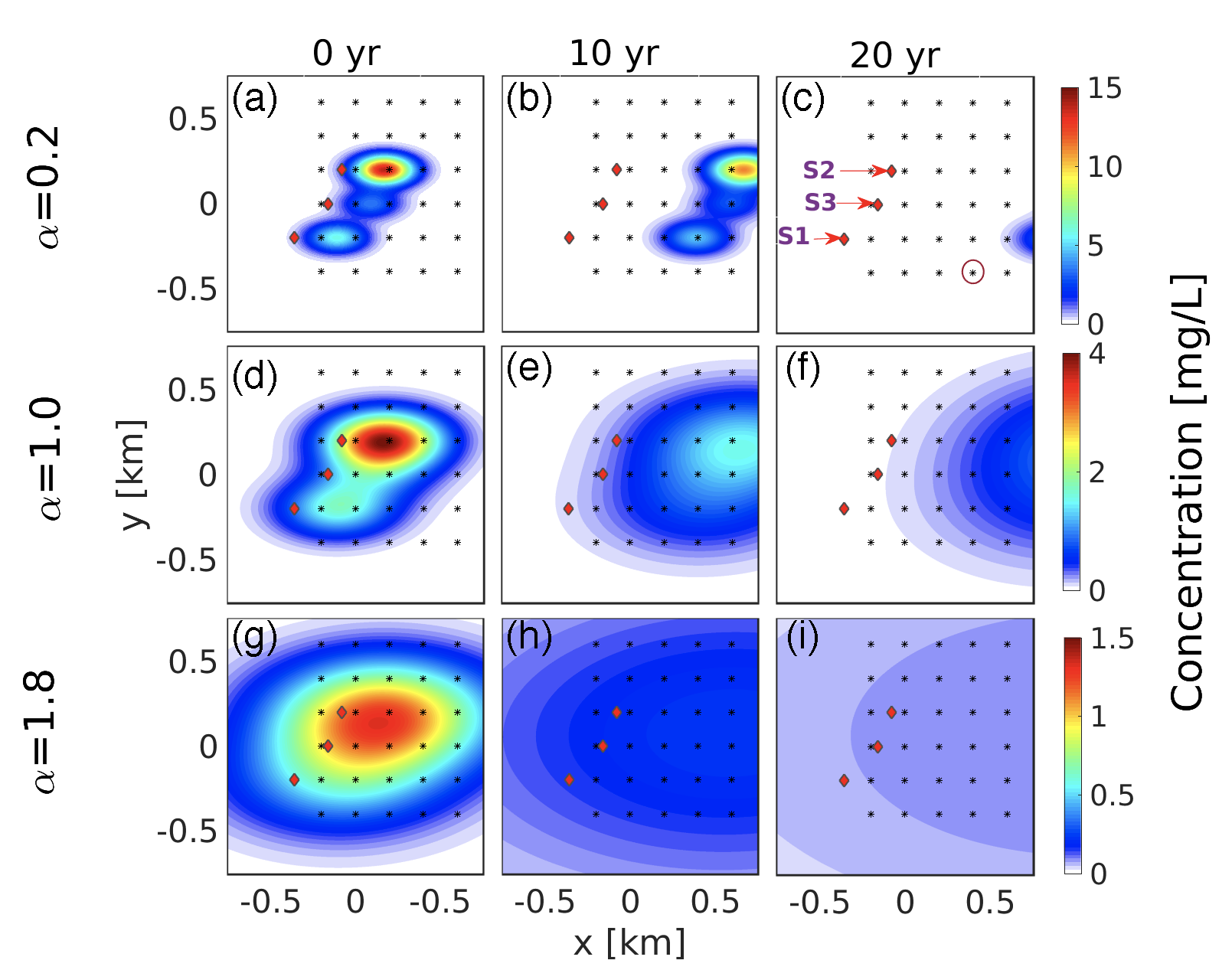}
 \caption{The configuration of point sources (red diamonds) and the array of detectors (black points). The panels in each row correspond to different values of the diffusion exponent, $\alpha$ (as specified), while the columns correspond to the denoted times. All the sources emitted at time $t_0=-5$ years. The source strengths are: $q_{1,2,3}=0.25, 0.6, 0.15\times10^{12}~\text{mg/km}$, respectively and their positions are (in $\text{km}$): $[-0.36,-0.2]$, $[-0.08,0.2]$ and $[-0.16,0]$, respectively. The color denotes the concentration at the different locations and times for each type of diffusion.}
 \label{fig:Conf}
\end{figure*}

\begin{figure*}[ht]
 \includegraphics[width=\linewidth,trim={2.5cm 0cm 5cm 1cm}, clip]{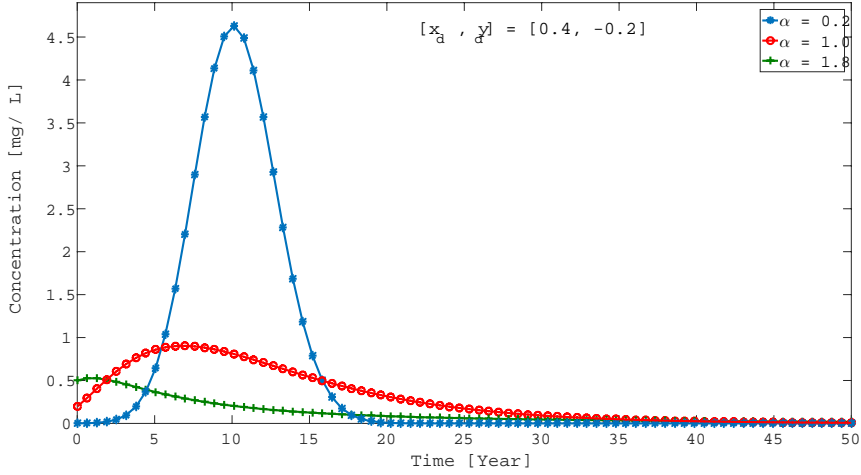}
 \caption{The concentrations recorded by a detector for three different types of diffusion as denoted by the values of the diffusion exponent, $\alpha$. The shown time series correspond to concentration at the circled detector in the top rightmost panel of Fig.~\ref{fig:Conf} at position $(x_d, y_d) = (0.4~\text{km}, -0.2~\text{km})$. All three sources ($S_1,S_2$, and $S_3$ marked by red diamonds in Fig.~\ref{fig:Conf}) release at $t_0= -5$ years.}
 \label{fig:detsigns}
\end{figure*}

To illustrate the complexity of the problem, we present in Fig.~\ref{fig:Conf} temporal snapshots of concentration profiles for subdiffusive ($\alpha =0.2$), normal diffusive ($\alpha =1$), and superdiffusive ($\alpha =1.8$) transport (to generate the synthetic data, we used parameters that are typical of the transport and dispersion of contaminants in an aquifer \cite{Boiandiff}). In each case,  emission originates instantaneously at $t_0= -5~\text{years}$ from three point sources, $S_1, S_2$, and $S_3$, marked by red diamonds. In all three cases, the flow is subjected to a drift velocity ${\bf{u}} = (u_x, 0)$, where $u_x=0.05~\text{km/year}$ in the $x$-direction. In the subdiffusive case, the contribution from each source remains relatively distinct throughout the illustrated time period of 20 years. However, for the normal and superdiffusive cases, the contributions from the three sources increasingly mix and become indistinguishable. In reality, complete spatio-temporal concentration profiles as shown in Fig.~\ref{fig:Conf} are unavailable. Concentration information is only available at a limited number of locations, where the detectors are positioned, shown as black points in Fig.~\ref{fig:Conf}. Examples of concentration times series from a single detector located at $(x_d,y_d) =(0.4~\text{km}, -0.2~\text{km})$ (marked by a red circle in the upper right panel of Fig.~\ref{fig:Conf}) are given in Fig.~\ref{fig:detsigns} for the three different types of diffusion discussed in Fig.~\ref{fig:Conf}. Again, as a result of the slow diffusion in the subdiffusive case, $\alpha=0.2$, the signal arrives at the detector in the form of a narrow and concentrated peak. However, for normal diffusion, $\alpha=1$, and super diffusion, $\alpha=1.8$, the signal is less concentrated and spread over a much longer time period as a result of the faster dispersion processes and the mixing of the source contributions.
The extension of hNMF, we report here, correctly estimates the unknown number of point sources and evaluates the sources' locations and strengths, as well as the medium properties (diffusion exponent, anisotropic generalized diffusion coefficients, and drift), from a limited number of time series similar to those shown in Fig.~\ref{fig:detsigns}. More complicated scenarios, such as temporally varying emission rates and spatially extended sources, were also considered.

%\onecolumngrid

%\twocolumngrid

%\twocolumngrid

\section{Methods}
\label{sec:Methods}

\subsection{Fractional Brownian motion Green's function}\label{subs:fbm}
There are various mechanisms that lead to anomalous diffusion \cite{Bouchaud_1990, Metzler_2000}. The simplest mechanism is fractional Brownian motion (fBm). fBm relies on a memory effect that can be introduced into the Langevin equation describing the dynamics of a particle \cite{mandelbrot1968fractional}. The mechanism was suggested to be relevant to many processes \cite{Biagini_2008}.
Despite the memory effects, fBm is a Gaussian process, in which, unlike classical Brownian motion, the increments of the fBm are not independent. The 2D Fokker-Planck equation describing Galilei invariant fBm is:
\begin{eqnarray}\label{gen_eq_fbm}
\frac{\partial C}{\partial t} =  \alpha t^{\alpha-1} \bigg(D_x \frac{\partial^2 C}{\partial x^2} + D_y \frac{\partial^2 C}{\partial y^2}\bigg) - u_x \frac{\partial C}{\partial x} + Q,
\end{eqnarray}
where $C$ is the concentration, $\alpha$ is the diffusion exponent ($0<\alpha<2$), $D_x, D_y$ are the generalized diffusion coefficients in the $x$ and $y$ directions, respectively, $u_x$ is the drift velocity in the $x$ direction (the coordinate system is chosen such that $u_y  \equiv 0$), and $Q(\vec{r},t)$ is the source function.

The Green's function (propagator) of the above Fokker-Planck equation, with a physical boundary condition requesting  the concentration to vanish at an infinite distance from the source, is given by (see \cite{metzler2004restaurant} for the 1D propagator):
\begin{eqnarray}\label{eq:Green2}
G_{\rm{fBm}}({\vec{r}},t) = \frac{1}{4 \pi \sqrt{D_x D_y} t^\alpha} e^{-\frac{(x - u_x t)^2}{4 D_x t^\alpha}}e^{-\frac{y ^2}{4 D_y t^\alpha}}.
\end{eqnarray}

For $\tilde{N}_s$ instantaneous point sources (i.e., $Q(\vec{r},t)=\sum\displaylimits_{s=1}^{\tilde{N}_s}q_s\delta(\vec{r}-\vec{r}_s)\delta(t-t_s)$), the concentration is given by the sum of the contributions of the different sources (the contribution of each point source is give by Eq. (\ref{eq:Green2}) multiplied by the source strength, $q_s$):
\begin{align}\label{eq:propNs}
&C\left(\vec{r},t\right) = \\
&\sum\displaylimits_{s=1}^{\tilde{N}_s} \frac{q_s}{4 \pi \sqrt{D_x D_y} \left(t-t_s\right)^\alpha}
e^{-\frac{(x - u_x \left(t-t_s\right)-x_s)^2}{4 D_x \left(t-t_s\right)^\alpha}}
e^{-\frac{\left(y-y_s\right)^2}{4 D_y \left(t-t_s\right)^\alpha}}.\nonumber
\end{align}

For release sources whose emission is finite in duration, the  source function, $ Q_{T}(\vec{r},t)$, takes the form,
\begin{align}
 Q_{T}(\vec{r},t)=\sum\displaylimits_{s=1}^{\tilde{N}_s}q_s\delta(\vec{r}-\vec{r}_s)\Theta(t-t^{on}_s)\Theta (t^f_s-t),
\end{align}
where $\Theta(t) = 0$, if $t <0$, and  $\Theta(t) = 1$, if $t \geq 0$, while $t^{on}_s,t^f_s$ are the beginning and end times of emission by the $s$'th source, respectively. In this case, for $t>t^f_s>t^{on}_s$ for $s\in[1,\tilde{N}_s]$, the concentration is given by:
\begin{eqnarray}
C_{T}\left(\vec{r},t\right) &=& \int\displaylimits_{-\infty}^t G_{fBm}\left(\vec{r}-\vec{r}\hskip 2pt ',t-t'\right)
Q_{T}\left(\vec{r}\hskip 2pt ',t'\right)dt' \nonumber \\
&=& \sum\displaylimits_{s=1}^{\tilde{N}_s}q_s\int\displaylimits_{t^{on}_s}^{t^f_s} G_{fBm}\left(\vec{r}-\vec{r}_s,t-t'\right)dt'.
\label{eq:test}
\end{eqnarray}
Here the limits of integration in Eq. (\ref{eq:test}) arise from the finite emission duration of the sources.

For spatially extended rectangular sources, the source function, $Q_{S}(\vec{r},t)$, takes the form,
%\begin{widetext}
\begin{align}
Q_{S}(\vec{r},t)=\sum\displaylimits_{s=1}^{\tilde{N}_s}q_s\Theta(x-x^l_s)\Theta(x^r_s-x)\Theta(y-y^l_s)\Theta(y^r_s-y)\delta(t-t_s).
\end{align}
Here, $x^l_s, x^r_s$ are the left/right boundaries of the $s$'th source and similarly for the $y$ coordinate to define rectangular sources. $t_s$ is time at which the $s$'th source emitted. The concentration in this case is given by:
\begin{align}
&C\left(\vec{r},t\right) = \int\displaylimits_{-\infty}^{\infty}\int\displaylimits_{-\infty}^{\infty} G_{fBm}\left(\vec{r}-\vec{r}\hskip 2pt ',t\right) Q_{S}\left(\vec{r}\hskip 2pt ',t\right)d\vec{r}\hskip 2pt ' = \sum_{s=1}^{\tilde{N}_{s}} q_s\int_{x^l_s}^{x^r_s}\int_{y^l_s}^{y^r_s} d{x'} d{y'} G_{fBm}(\vec{r}-\vec{r}\hskip 2pt ',t - t_i) =
\\
&\sum_{s=1}^{\tilde{N}_{s}} \frac{q_s}{4} \left[\mathrm{erf}\left(\frac{u_x(t-t_s)+x^r_s-x}{\sqrt{4D_x (t-t_s)^\alpha}}\right)-
\mathrm{erf}\left(\frac{u_x (t-t_s)+x^l_s-x}{\sqrt{4D_x (t-t_s)^\alpha}}\right)\right]\times
\left[\mathrm{erf}\left(\frac{y^r_s-y}{\sqrt{4D_y (t-t_s)^\alpha}}\right)-
\mathrm{erf}\left(\frac{y^l_s-y}{\sqrt{4D_y (t-t_s)^\alpha}}\right)\right] \nonumber.     
\end{align}
%\end{widetext}
\subsection{The Hybrid Nonnegative Matrix Factorization (hNMF) method}\label{sec:hNMF}
The Hybrid Nonnegative Matrix Factorization (hNMF) method reported in Ref. \cite{Boiandiff} combines: (i) Green's function of the Fokker-Planck equation (FPe) that describes normal diffusion transport with  (ii) a nonlinear iterative minimization procedure and (iii) a customized clustering, introduced previously as a method for estimating the number of the latent features in NMF \cite{2013alexandrovsignatures}. It was shown that the hNMF can successfully identify the unknown number of release sources, as well as the parameters of the FPe. To do this, the hNMF's algorithm explores the space of plausible solutions and narrows the set of possibilities by estimating the optimal number of release sources needed to reconstruct the observed data in a robust manner. Here, we generalize the hNMF to be adequate for fBm processes and show that the extended method can accurately determine the number of release sources and their locations and strengths, as well as the diffusion exponent and the transport properties of the medium ($D_x, D_y, u_x$). We also demonstrate the importance of this extension for the correct identification of the unknown number of release sources. In what follows, we assume point sources with the Green’s function given by Eq. \eqref{eq:Green2}; for the other source types, it has to be replaced by the appropriate Green’s function.
\subsubsection{Nonnegative Matrix Factorization (NMF)}
The usual interpretation of NMF is as a method for a low-rank matrix approximation of the observed data-matrix, $C$, whose size is $T\times n$, by two unknown matrices $W$ and $H$, $C\approx WH$, both containing one small dimension, $N_s$ (which is the estimate for the actual number of sources, $\tilde{N}_s$). This approximation (aka matrix decomposition) is performed through non-convex minimization with a given distance metric, $||...||_{dist}$: min$||C_{ij}-\sum^{N_s} _{s=1}W_{is}H_{sj}||_{dist}$ constrained by the non-negativity of $W$ and $H$, $W_{is}\geq0$; $H_{sj}\geq0$. NMF has proven very useful for face recognition, text recognition, dimension reduction, unsupervised learning, anomaly detection, Blind Source Separation (BSS), etc. \cite{cichocki2009nonnegative}. NMF is underpinned by a well-defined statistical model of superimposed components that, when the distance metrics $||...||_{dist}$ is the Euclidean distance, can be treated as a Gaussian mixture model \cite{fevotte2009nonnegative}. In this probabilistic interpretation, NMF is equivalent to the expectation-maximization (EM) algorithm \cite{dempster1977maximum}. EM is developed to find the maximum likelihood estimates of parameters in statistical models, when the model depends on unobserved, i.e., latent or hidden, variables.  In this probabilistic interpretation of NMF, the observables,  $c_1, c_2, ..., c_n$ ($c_i$ is a column vector with $T$ elements), which are the columns of the data, $C$, are generated by $N_s$ latent variables $h_1, h_2,..., h_{N_s}$. Specifically, each observable $c_i$ is generated from a probability distribution with mean,  $\langle c_i \rangle=\sum^{N_s} _{ s=1}W_{is}h_{s}$, where $N_s$ is the (unknown) number of latent variables. The influence of $h_s$ on $c_i$ is through the basis patterns/features of the considered phenomenon $w_{:s}$ represented by the columns of $W$ \cite{lee1999learning}. It is known that the probabilistic interpretation of NMF is particularly valuable when dealing with stochastic signals \cite{fevotte2009nonnegative}.

In our case, the observed data, $C$, are formed by the mixing of $\tilde{N}_s$ signals, at the locations of each one of the $n$ detectors, $ \vec{r}_d\equiv(x_d,y_d); d=1,...,n$ at the times of the records $t_m; m =1, 2, ..., T$. Each one of the $n$ detectors is situated at a different location, $\vec{r}_d = (x_d, y_d)$, and measures a mixture of the signals originating from $\tilde{N}_s$ point sources located at the positions $\vec{r}_s=(x_s, y_s), s=1, 2 ,..., \tilde{N}_s$ with respective strengths $q_s$. The fractional diffusion, characterized by the diffusion exponent $\alpha$, occurs in a medium with generalized diffusion coefficient ($D_x$, $D_y$) and drift $u_x$. The transient signal recorded by the $d^{th}$ detector is assumed to be generated from a normal probability distribution with mean
\begin{equation}\label{eq:BSScontaminant1}
\langle {c}_{\vec{r}_d,t_m}\rangle  = \sum^{N_s} _{s=1}{W}_{s,(\vec{r}_d,t_m)}{H}_{s} ,
\end{equation}
where the index $\vec{r}_d$ denotes the position of the recording detector, and the index $t_m$ denotes the time point when the observed data have been recorded. Further, $W_{s,(\vec{r}_d,t_m)}$ are the fBm Green's functions (Eq. (\ref{eq:Green2})) of each source. $W_{s,(\vec{r}_d,t_m)}$ in Eq. (\ref{eq:BSScontaminant1}) can be considered as a specific kernel \cite{zhang2006non} that depends on space-time. $H$ contains $N_s$ hidden variables, independent of time and space, that generate the observables $c(\vec{r}_d, t_m)$ at points $(\vec{r}_d,t_m)$ via the weights $W_{s,(\vec{r}_d,t_m)}$. Note that both $W_{s,(\vec{r}_d,t_m)}$ and $H_s\equiv q_s$ are non-negative.

For given detectors' locations, time points, and the functional form of the fBm Green's function (Eq. (\ref{eq:Green2})), our aim is to estimate the (unknown
) number of sources, $\tilde{N}_s$ and to determine the source parameters, $x_s$, $y_s$, $q_s$ ($s\in[1,\tilde{N}_s]$) and the medium transport characteristics $u_x$, $D_x$, $D_y$, as well as the diffusion exponent $\alpha$. In this interpretation and from Eq. (\ref{eq:BSScontaminant1}), the minimization can be performed by nonlinear least squares on the objective function $O$,

\begin{equation}
\label{eq:NMFof}
{O} = \sum_{d=1}^n\sum_{m=1}^{T} \left(C_{\vec{r}_d, t_m} -\sum_{s=1}^{N_s} W_{s,(\vec{r}_d, t_m)} H_{s} \right)^2,
\end{equation}
where $d$ marks the detectors, and $m$ marks the time points of the records.

The minimization of the objective  function, $O$, assumes that each measurement at a given space-time point $(\vec{r}_d, t_m)$ is an independent Gaussian-distributed random variable. If each detector has its own distinct
(possibly time-dependent) measurement error, the objective function in (\ref{eq:NMFof}) should be replaced by a weighted sum of least squares, where each deviation from the corresponding observation is weighted by the inverse square of its measurement error.

\subsubsection{Custom clustering and NMFk}\label{subsubs:clustering}

 NMF is sufficient to carry a constrained optimization problem to extract desired parameters when the number of sources, $\tilde{N}_s$, is known; however, we rely on the signals measured by the detectors, which record mixtures arising from an unknown number of sources. To determine this unknown number of sources (that is, the number of the latent features), we utilize an approach called NMFk, introduced in earlier works \cite{2013alexandrovsignatures, boianBSS, Boiandiff}.

NMFk explores the possible number of sources, $\mathcal{N}_s$, starting from $\mathcal{N}_s = 1, 2, ..., P$ ($P$ is less than min($n$, $T$)). For each explored number of sources, $\mathcal{N}_s$, a set $U_{\mathcal{N}_s}$ (we call it a run) of $M\sim200$ minimizations
are computed, each minimization generated from random initial guesses for the unknown parameters (within specific physical bounds). The set of $M$ solutions in the set $U_{\mathcal{N}_s}$, for $\mathcal{N}_s$ sources, is given by

\begin{equation}
\begin{split}
{ U }_{ \mathcal{N}_s }=([(X, Y, Q)_1, (u_x, D_x,D_y,\alpha)_1], ...\\, [(X, Y, Q)_M, (u_x, D_x,D_y,\alpha)_M]),
\end{split}
\end{equation}

where $(X, Y, Q)_i $ denotes the unknown coordinates and strengths of the $\mathcal{N}_s$ sources; $ [(x_1, y_1, q_1), ...,  (x_{\mathcal{N}_s}, y_{N_s}, q_{\mathcal{N}_s})]_i$ in the $i^{th}$ NMF minimization with advection velocity, generalized diffusion coefficients and  diffusion exponent $(u_x, D_x,D_y,\alpha)_i$. After we build the set of $M$ NMF solutions (for $\mathcal{N}_s$ sources), the parameters pertaining to sources $ [(x_1, y_1, q_1), ...,  (x_{\mathcal{N}_s}, y_{\mathcal{N}_s}, q_{\mathcal{N}_s})]_i$ are subjected to customized clustering into $\mathcal{N}_s$ number of clusters. For a single NMF run with $\mathcal{N}_s$ sources, a total of $M\sim 200$ simulations are carried out, which gives $M$ tuples $[(X, Y, Q)_i, (u_x, D_x,D_y,\alpha)_i]$.  Each of the tuples, $[(X, Y, Q)_i, (u_x, D_x,D_y,\alpha)_i]$, represents a distinct solution for nominally equivalent NMF minimizations, where the difference arises as a result of the random initial guesses. Next, we perform customized clustering, assigning the parameters of each $W_{s,(\vec{r}_d,t_m)}$ of all $M$ solutions to one of $\mathcal{N}_s$ clusters. This customized clustering is similar to k-means clustering but with an additional constraint which constrains the number of elements in each of the clusters to be equal to the number of solutions $M$. For example, with $M=200$, each one of the $\mathcal{N}_s$ identified clusters must contain exactly 200 solutions. This condition has to be enforced since each NMF minimization (producing a given $[(X, Y, Q)_i, (u_x, D_x,D_y,\alpha)_i]$ tuple) contributes only one solution, and accordingly has to supply exactly one element to each cluster. During the clustering, similarity between the elements is measured by cosine similarity, which calculates the cosine of the angle between two vectors and is a natural choice for similarity between non-negative vectors \cite{korenius2007principal}.

Further, to identify the optimum number of sources, NMFk calculates the clusters' stability for each explored number of sources, $\mathcal{N}_s$. The optimum number of sources, $N_s$, which is used to estimate the true $\tilde{N}_s$, is the number of sources for which the corresponding clusters are relatively stable and separable and whose centroids result in a small reconstruction error (see below).

To quantify the stability and separability of the clustering for a given number of sources, NMFk utilizes the Silhouette statistics, $S$ \cite{rousseeuw1987silhouettes}, which is developed to measure the similarity between an element and the elements of its own cluster compared to the centroids of other clusters. The $S$ values are between $[-1, 1]$, and $S$ measures how well each element has been classified by the clustering.
The main idea of NMFk is to use the cohesion of the clusters (how compact they are) and the separation between them as a measure of the stability of the solutions of the minimization with different random initial guesses and hence the quality of a particular choice of $\mathcal{N}_s$. In previous works, NMFk even "shuffled" randomly the observed data, $C$,  to increase the effect of robustness of the average solutions \cite{alexandrov2013deciphering,stanev2018unsupervised} and for the large scale variations~\cite{chennupatidistributed}.

The intuitive reasoning behind the idea of stability is as follows. In the case of underfitting, i.e., for solutions with a number of sources less than the actual
number of sources,  the clustering could be good; for example, several of the sources could be combined to produce a "super cluster." However, the
clustering will break down significantly in the case of overfitting, when the estimated number of sources exceeds the true number of sources. Indeed, in this case, we do not expect the solutions to be well clustered, since there is no unique way to reconstruct
the solutions with number of clusters $> \tilde{N}_s$, and at least some of the clusters will be artificial, rather than real entities.

The other metrics that NMFk utilizes is the relative reconstruction error, $R = ||C-WH ||/||C||$, which measures the relative deviation of the obtained solution, $WH$, from the original data, $C$.  The reconstruction error, $R$, evaluates the accuracy with which the average solutions,
that is, the solution constructed with the parameters taken from the centroids of the clusters, reproduce the observed data, $C$. In general, the solution accuracy increases (while the stability of the clusters decreases) with the increase in the number of the sources (increasing the number of the parameters in the minimization).

In summary, NMFk calculates the average Silhouette width, $S$, and
the average reconstruction error, $R$, for each choice of the unknown number of sources, in order
to estimate the true number of release sources, $\tilde{N}_s$. NMFk determines $\tilde{N}_s$ to be equal to the number of sources that accurately reconstruct
the observations (i.e., their relative reconstruction error, $R$, is small enough) and the clustering of the sets of solutions corresponding to $N_s$, obtained with random initial conditions, to be sufficiently robust (i.e., the average Silhouette width, $S$, to be close to 1).

\section{Generation of Synthetic Data}\label{sec:sd}
In order to validate the generalization of the hNMF method, we generated synthetic data sets using Eq. (\ref{eq:propNs}) for different types of diffusion (dictated by the values of $\alpha$). The parameters that were used here to generate the synthetic data are typical for groundwater contamination \cite{Boiandiff}. The parameters correspond to the case of normal diffusion, and for the case of anomalous diffusion, the values of the Brownian dispersion coefficients were used as generalized diffusion coefficients. We used: $D_x = 0.005 ~\text{km}^2/\text{year}^{\alpha}$ and $D_y = 0.00125 ~\text{km}^2/\text{year}^{\alpha}$ , $t_0 = -5 ~\text{years}$ and $\alpha\in [0,2]$ ($\alpha<1$ corresponds to subdiffusion and $\alpha>1$ corresponds to superdiffusion). The choices of source and detector positions that were used for the results in Figs. \ref{fig:fit} and \ref{fig:silrec} are shown in Fig.~\ref{fig:Conf}. In Fig. \ref{fig:experiments}, two different configurations are shown. Each synthetic data set contains 30 time series for $0 ~\text{years} < t < 50~\text{years}$ with equal sampling intervals such that each series includes a total of 80 concentration values (see Figs.~\ref{fig:detsigns} and \ref{fig:experiments} for examples).
\section{Numerical Experiments}

\subsection{Generalized hNMF is needed for the accurate identification of fBm sources}
\begin{figure*}[ht]
 \includegraphics[width=\linewidth]{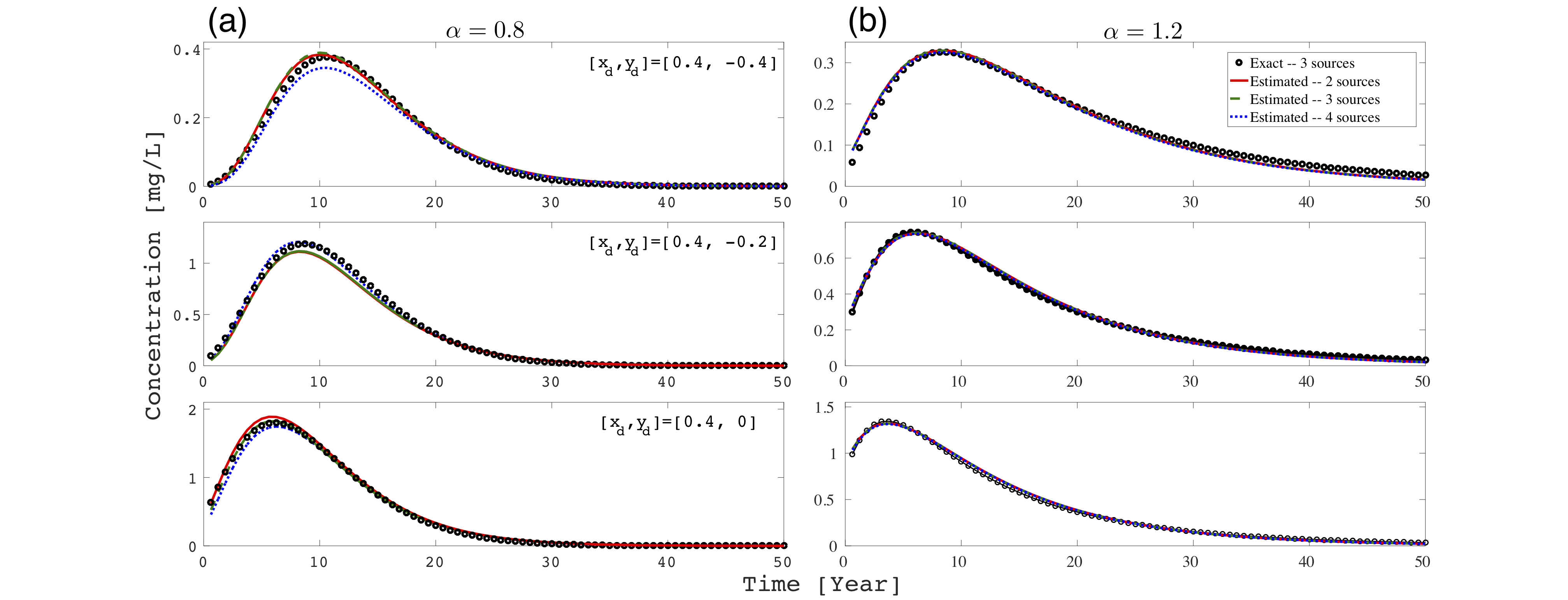}
 \caption{Measured (black lines) and estimated signals (green, red, and blue markers) at three detectors. The left column[panel (a)] shows the reconstructed signal for 2, 3 or 4 sources and $\alpha=0.8$, and the right column [panel (b)] shows the same information for $\alpha=1.2$. The different rows correspond to different detectors whose coordinates are specified. The estimated signals are all for normal diffusion $\alpha=1$.}
  \label{fig:fit}
\end{figure*}

In order to illustrate the importance of the generalized hNMF method, we considered synthetic data generated using $\alpha=0.8$ (subdiffusion) and $\alpha=1.2$ (superdiffusion), the parameters listed in section \ref{sec:sd}, and the configuration of sources and detectors shown in Fig. \ref{fig:Conf}. Both values of the diffusion exponent are close to normal diffusion, $\alpha=1$.
We ran the hNMF with $\alpha$ set to 1 (i.e., removing the degree of freedom of the diffusion exponent).
The reconstructed signals at the locations of three different detectors are depicted in Fig. (\ref{fig:fit}).
The synthetic data corresponding to $\alpha=0.8$ and $\alpha=1.2$ are presented by  solid black lines in the left(right) panels, and the circles represent the reconstructed signals at these locations using 2, 3, or 4 point sources.  Fig. (\ref{fig:fit}) demonstrates that in all cases, the detector observations are well approximated despite the wrong propagator used and the wrong number of sources. It is, therefore, clear that the ability to approximate detector observations alone is not sufficient to determine the number of sources and their locations or the diffusion characteristics. If the $\alpha$ is not forced, we see slightly better reconstructions as generalized hNMF recognizes the correct number of sources with the appropriate diffusion characteristics. 
%
%\onecolumngrid

\begin{figure*}[ht]
 \includegraphics[width=0.85\linewidth]{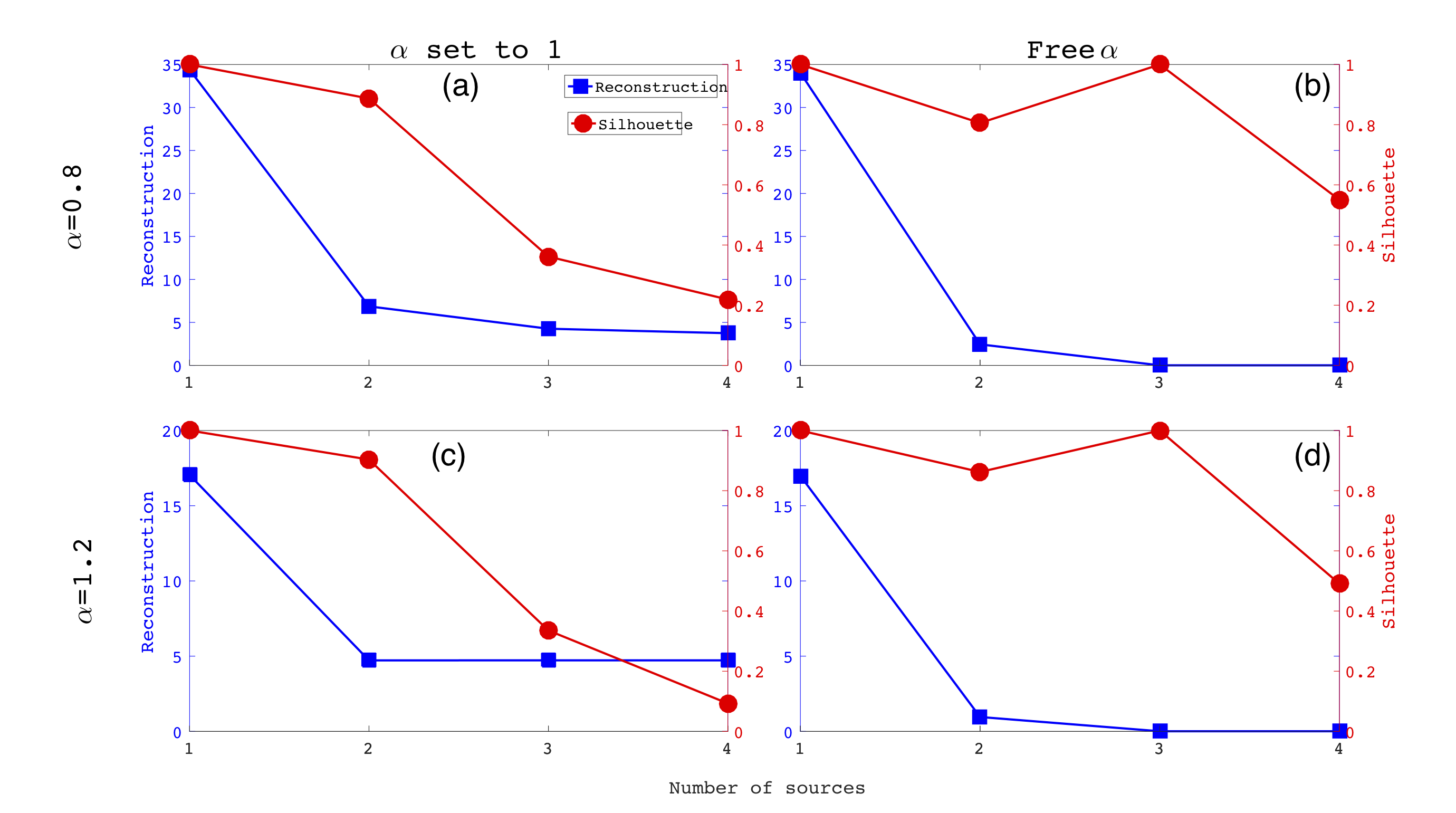}
 \caption{The reconstruction and silhouette measures for different numbers of sources. The upper row[(a),(b)] corresponds to $\alpha=0.8$ and the lower row[(c),(d)] to $\alpha=1.2$. The left column corresponds to the method with $\alpha$ set equal to one (i.e., looking for the best normal diffusion parameters to describe the measurements), and the right column corresponds to the full method with $\alpha$ being extracted by the method.}
 \label{fig:silrec}
\end{figure*}

%
%\twocolumngrid
In Fig.~\ref{fig:silrec}, we present the  reconstruction metrics, $R$, and average Silhouette, $S$, for the different numbers of sources ($\mathcal{N}_s$) for $\alpha=0.8$ (subdiffusion) (upper row) and $\alpha=1.2$ (superdiffusion) (lower row). The left column corresponds to a forced $\alpha=1$, that is, to the old hNMF. The right column corresponds to the results of the generalized hNMF with the diffusion exponent derived by the minimization described in section \ref{sec:hNMF}.
From Fig.~\ref{fig:silrec}, it can be concluded that with  $\alpha$ set to one, (left column), both the reconstruction, $R$, and the average Silhouette, $S$, decrease as the number of sources increases, for both subdiffusion and superdiffusion. This observation implies that, despite the better fit of the data by the estimated parameters, the increase in the number of sources worsens the clustering quality (a reduced Silhouette). The reduction of the average Silhouette stems from the fact that the old hNMF model used to describe the data does not account for anomalous diffusion.
In the right column, we show that when the generalized hNMF is used, with the diffusion exponent extracted by the minimization, for both subdiffusion and superdiffusion, the maximum value of the silhouette and the minimal value of the reconstruction are obtained for the correct number of sources (three) that was used to generate the synthetic data.
%
%\onecolumngrid

%
%\twocolumngrid
\subsection{Generalized hNMF  accurately identifies fBm sources and properties of the medium}
\begin{figure*}[ht]
 \includegraphics[width=\linewidth]{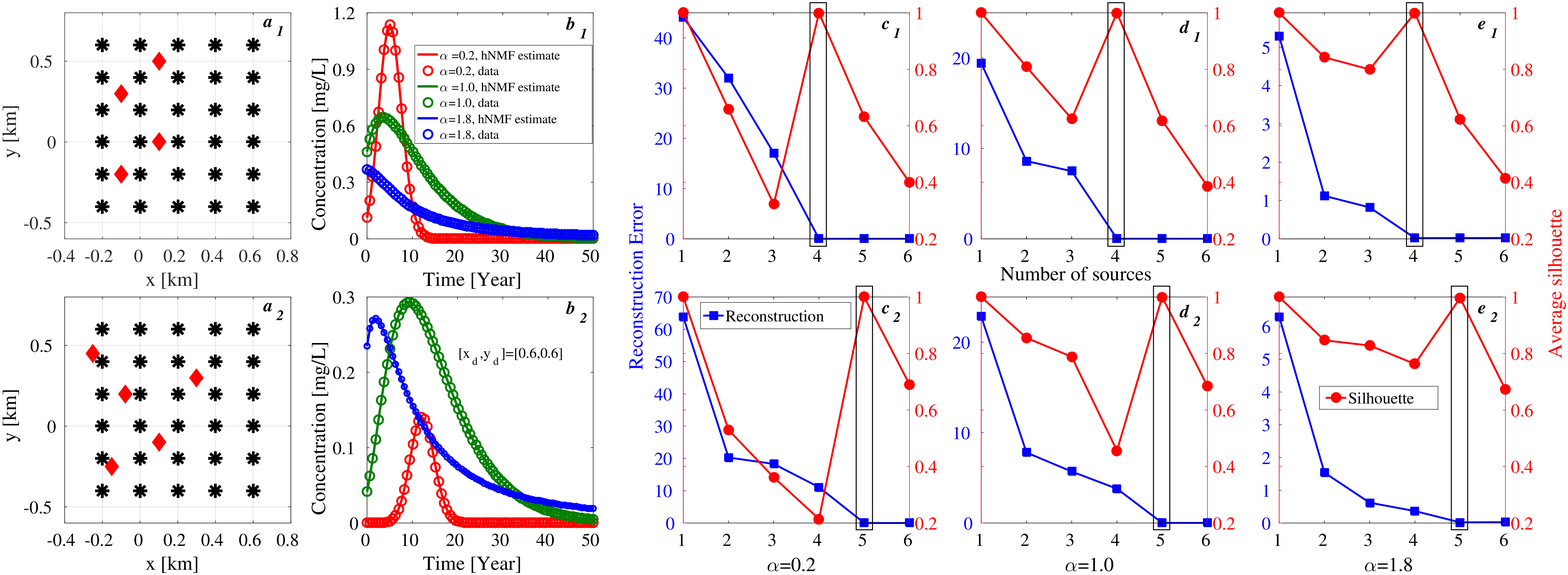}
 \caption{Generalized hNMF benchmarks for two spatial configurations with four (upper panel 1) and five (bottom panel 2) release sources. Subfigures, noted by $a_i$, $b_i$, $c_i$, and $e_i$; $i=1,2$, represent ($a_i$) spatial configuration of sources and detectors, ($b_i$) time series of the concentration measured by a detector at $\vec{r}_d$ = (0.6, 0.6) with the reconstructed signal; solid line represents recorded signals, and reconstructed signals are plotted with red markers ($\alpha =0.2$, subdiffusion), green markers ($\alpha =1.0$, normal diffusion), and blue markers ($\alpha=1.8$, superdiffusion). Subfigures, ($c_i$), ($d_i$) and ($e_i$) represent the reconstruction error, $R$, (left, blue) and the average Silhouette width, $S$, (right, red) of the solutions; the shadowed box depicts the true number of sources, $\tilde{N}_s$, which is equal to the optimal number of sources determined by the hNMF.}
  \label{fig:experiments}
\end{figure*}

We applied the generalized hNMF method to different configurations of point sources and detectors and found that the method provides a correct estimate of the number of sources; their strengths and positions; and medium characteristics (drift velocity, generalized diffusion coefficients, and the diffusion exponent).
Figure \ref{fig:experiments} demonstrates the results of our extended hNMF method, for two different numbers of release point sources and spatial configurations, and for three values of the diffusion exponent corresponding to subdiffusion, normal diffusion, and superdiffusion.  Subfigures $b_1$ and $b_2$ show the reconstruction of the original signal measured by the detector located at (0.6,0.6). The solid line represents the original signal recorded by the detector, and the markers are the solution estimates by the extended hNMF method. It is clear that the signals are well reconstructed for all cases of the diffusion exponent $\alpha$. Subfigures c, d, and e demonstrate that the extended hNMF identified the correct number of release sources.\\
%\onecolumngrid

%\twocolumngrid
The subfigures c, d and e of the two panels show the average reconstruction errors (left, blue) and average Silhouette widths (right, red) used to determine the unknown number of release sources present in the system. It is clear from the subfigures that our extended method correctly estimates the number of sources, their characteristics, and the transport characteristics for all the cases of the diffusion exponent considered here.

We also tested the performance of the method with configurations consisting of temporally extended and spatially extended sources. In all these cases, the generalized hNMF method provides accurate estimates of the duration of emission and the spatial extension of the sources. We also verified that for the synthetic data that was generated using point sources, when the method was implemented using the propagator for spatially extended sources, the estimated size of the sources was smaller than any spatial scale of the configuration (the dispersion radius between sequential measurements, the distance between detectors, the distance between sources, and the drift distance between sequential measurements).

\subsection{Noisy data}

Real data often includes noise. The noise can be due to the measuring device/s or due to other effects which are not accounted for in the Green function describing the process. In many realistic cases, the noise is proportional to the signal. In order to test the applicability and robustness of the hNMF, we considered noisy data. The data was generated in the same way described in section \ref{sec:sd} for the configuration shown in Fig. \ref{fig:Conf}. In order to add the noise, the precisely calculated concentration (at each time and position) was multiplied by a random number drawn from a uniform distribution according to:

\begin{equation}\label{eq:noise}
 \mathcal{C}\left(\vec{r},t\right)=C\left(\vec{r},t\right)\left(1+\mu z\left(\vec{r},t\right)\right).
\end{equation}

The probability density function of $z\left(\vec{r},t\right)$ is simply
\begin{equation}\label{eq:noisepdf}
 p\left(z\left(\vec{r},t\right)\right)=
 \begin{cases}
  1/2\ \ \ \ \ -1\leq z\leq 1,\\
  0\ \ \ \ \ \ \ \ \ else.
 \end{cases}
\end{equation}

\begin{figure*}[ht]
 \includegraphics[width=\linewidth]{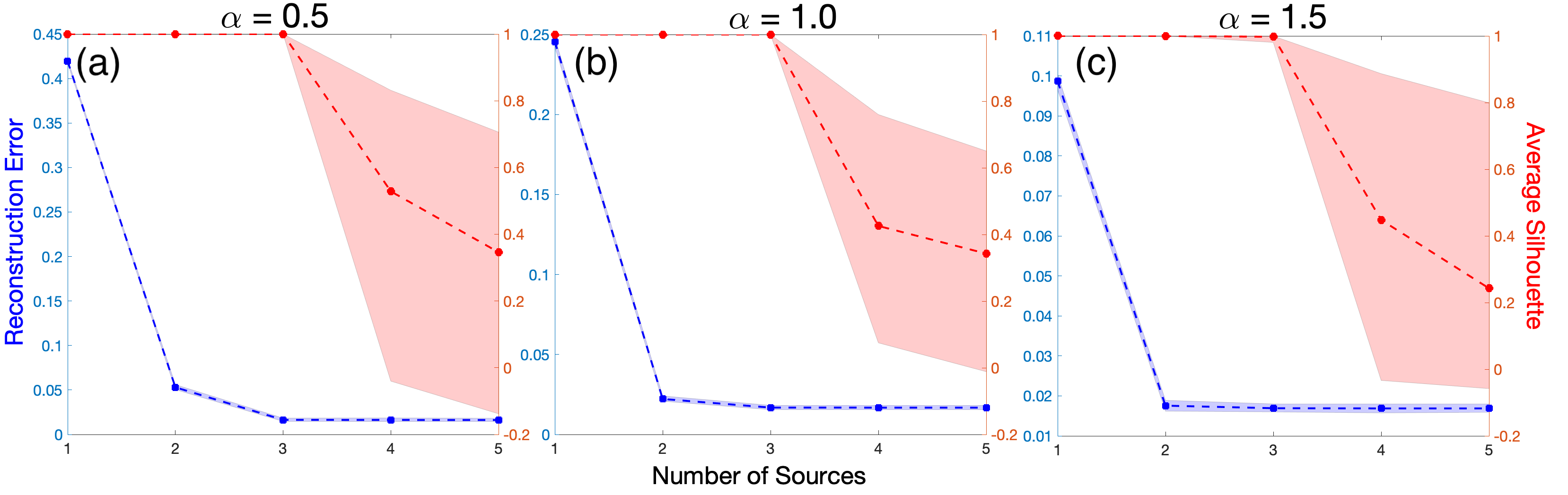}
 \caption{Reconstruction error and average silhouette for noisy data. The three panels correspond to the denoted values of $\alpha$. The configuration is the same as shown in Fig. \ref{fig:Conf} and the noise amplitude is $\mu=0.03$ (see eq. \eqref{eq:noise}). The shaded area represents the range of values obtained from 50 realizations of the noise.}
  \label{fig:noisesr}
\end{figure*}

In Fig. \ref{fig:noisesr}, we present the reconstruction error and the average silhouette for different numbers of sources. The three panels correspond to the specified values of the diffusion exponent, $\alpha$ (corresponding to sub, normal and super diffusion).
%
%\onecolumngrid

%\twocolumngrid
For the cases of subdiffusion ($\alpha=0.5$) and normal diffusion ($\alpha=1$), the reconstruction error and the average silhouette show a similar behavior to that found for data in the absence of noise, and the correct number of sources is identified. For the superdiffusive case, where the mixing of the signals from the three sources is stronger, for some realizations of the noise, the average silhouette does not show the same sharp decrease, and the identification of the number of sources is more challenging. For lower values of the noise, the hNMF shows the same performance as for the exact data.

\begin{figure*}[ht]
 \includegraphics[width=\linewidth]{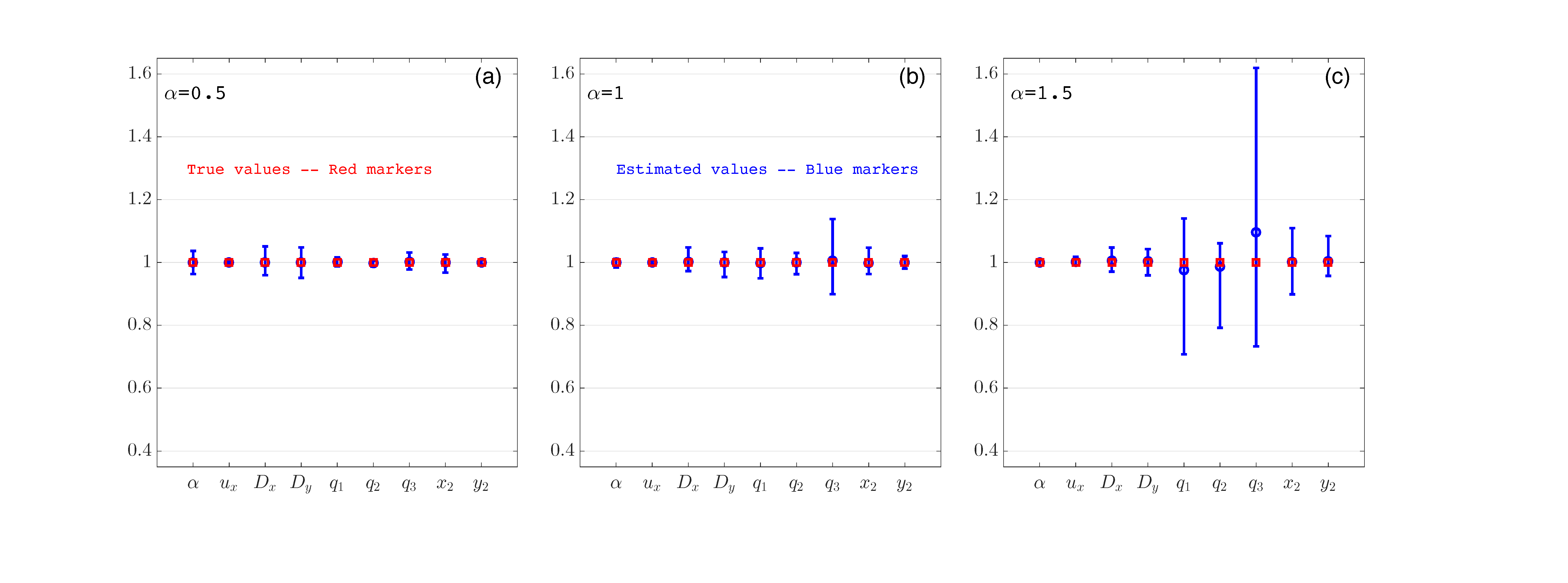}
 \caption{The normalized hNMF estimated parameters and the associated uncertainties. The three panels correspond to the denoted values of $\alpha$. The configuration is the same as shown in Fig. \ref{fig:Conf} and the noise amplitude is $\mu=0.03$ (see eq. \eqref{eq:noise}). The error bars represent the range of values obtained from 50 realizations of the noise.}
  \label{fig:noisep}
\end{figure*}

The noise also introduces uncertainty in the estimated values of the parameters. In Fig. \ref{fig:noisep}, we present the estimated values of the parameters for the same diffusion exponents shown in Fig. \ref{fig:noisesr}.
%\onecolumngrid

%\twocolumngrid
Each parameter was normalized by the true value in order to present the relative error and uncertainty.
The uncertainty range is larger for larger diffusion exponents where the mixing of the signals from the different sources is larger. For the normal diffusion and the superdiffusion, the uncertainty in the estimation of the amplitude of the third source is the largest. The noise is proportional to the total signal, and the third source has the smallest amplitude. Therefore, the noise relative to the signal from this source is much larger and it affects the uncertainty associated with its estimated amplitude.

\section{Conclusions}
Inverse modeling of anomalous diffusion, is of great interest in many fields, including material science, physics, finances, biology, and many others.
The value of an inverse modeling approach lies in its ability to correctly estimate the unknown number of release sources and their characteristics, as well the medium characteristics and the diffusion exponent.
We demonstrated that the generalized hNMF introduced here, which is based on the integration of unsupervised learning and Green's function of the fBm governing equation, is capable of providing accurate estimates of the number of sources, their properties, and the medium characteristics that result in sub, Brownian, or super diffusion. The method was shown to work with different types of sources, including point sources, temporally extended sources, and spatially extended ones with various configurations.
The generalized hNMF method also provides information regarding the validity of the type of the propagator that is used. If the propagator is not appropriate (e.g., if it is the propagator of normal diffusion), the method showed that it is impossible to have maximal Silhouette and minimal reconstruction metrics for any number of sources. When the adequate propagator is used, the optima that were found corresponded to the configurations that were used to generate the synthetic data.
The results presented here and previous successful applications of the hNMF for inverse problems suggest that this method might be further extended to account for different mechanisms of anomalous diffusion (such as the continuous time random walk and quenched disorder). Of particular interest are generalizations of the hNMF that would enable the extraction of the scaling relations based on limited observations.

\section{Acknowledgments}
This research was funded  by LANL LDRD grant 20190020DR, and resources were provided by the Los Alamos National Laboratory Institutional Computing Program, supported by the U.S. Department of Energy National Nuclear Security Administration under Contract No. 89233218CNA000001. Data sharing is not applicable to this article as no new data were created or analysed in this study.

\bibliographystyle{abbrv}
\bibliography{Main_Diff}%

\end{document}